\documentclass[conference]{IEEEtran}
\IEEEoverridecommandlockouts
\usepackage{cite}
\usepackage{amsmath,amssymb,amsfonts}
\usepackage{algorithmic}
\usepackage{graphicx}
\usepackage{textcomp}
\usepackage{xcolor}
\usepackage{tabularx}
\usepackage{multirow}
\usepackage{booktabs}
\usepackage[hidelinks]{hyperref}
\usepackage{listings}
\usepackage{float}
\floatstyle{plain} % optionally change the style of the new float
\newfloat{Code}{H}{myc}
\def\BibTeX{{\rm B\kern-.05em{\sc i\kern-.025em b}\kern-.08em
    T\kern-.1667em\lower.7ex\hbox{E}\kern-.125emX}}
\definecolor{bg}{rgb}{0.95,0.95,0.95}

\def\framework{elasticAI.explorer}

\begin{document}
\title{\framework:\\Towards a Unified End-to-End Framework for\\Hardware-Aware Neural Architecture Search

}
\author{
    \IEEEauthorblockA{
    Natalie~Maman, Florian~Hettstedt, Andreas~Erbslöh, Gregor~Schiele\\
    \textit{Intelligent Embedded Systems Lab, University of Duisburg-Essen},\\
    \textit{Faculty of Computer Science, Duisburg, Germany} \\
    \{first-name\}.\{last-name\}@uni-due.de
    }
}

\maketitle

%%%%%%%%%%%%%%%%%%%%%%%%%%%%%%%%%%%%%%%%%%%%%%%%%%%%%%%%%%%%%%%%%%%
%% BODY
%%%%%%%%%%%%%%%%%%%%%%%%%%%%%%%%%%%%%%%%%%%%%%%%%%%%%%%%%%%%%%%%%%%
\begin{abstract}
Neural Architecture Search (NAS) has become an important approach for automatically designing neural networks under task-specific and hardware-specific constraints. However, many existing NAS frameworks tightly couple search space definitions, model implementations, and deployment pipelines, making extension to new hardware platforms and custom operators difficult. In this paper, we present the elasticAI.explorer, an extensible Python framework for hardware-aware NAS built on top of Optuna. The framework introduces a YAML-based search space specification that dynamically translates into executable neural network models during sampling. The approach supports layer-wise, cell-based, and hierarchical search spaces while maintaining a unified interface for optimization and deployment. Beyond architecture generation, the framework integrates hardware-specific code generation, Docker-based cross-compilation toolchains, and automated creation of on-device benchmarking binaries, enabling hardware-in-the-loop NAS workflows. The system further provides extensible evaluators for FLOPs, parameter count, and latency estimation. The \framework~aims to reduce the engineering overhead of embedded AI deployment and accelerate research on hardware-aware NAS for heterogeneous accelerator platforms.
\end{abstract}
\vspace*{-5mm}
\section{Introduction}
\begin{figure*}[t]
    \centering
    \includegraphics[width=0.78\textwidth]{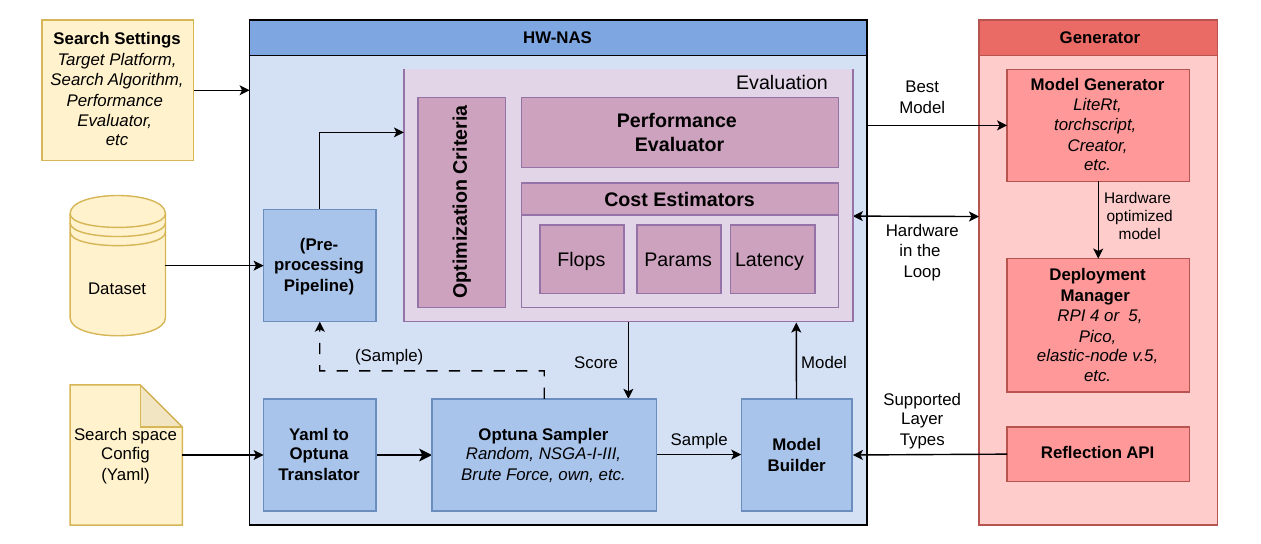}
    \caption{Overview of the end-to-end architecture within the \textit{\framework} from defining the search space over candidate sampling to the deployment. Interpretation of the highlighted boxes: yellow = user inputs, blue = search space sampling and model building, purple = model performance evaluation / estimation, red = hardware deployment part}
    \label{fig:overview}
\end{figure*}
Designing neural networks for embedded deep learning systems requires navigating a large space of architectural choices under task-specific and hardware-specific constraints. Neural Architecture Search (NAS) can automate parts of this process by exploring candidate architectures and identifying suitable trade-offs between predictive performance and deployment efficiency~\cite{elsken2019survey}. However, NAS experiments are often difficult to reuse and compare when search spaces are defined directly in code or tailored to a specific model family.

For embedded hardware-aware NAS, this challenge is coupled with deployment concerns. Candidate architectures must not only achieve high predictive performance, but also satisfy hardware constraints such as latency, memory footprint, resource usage, and compatibility with target-specific toolchains. Consequently, a practical framework should provide extensible abstractions that connect search space specification, model generation, deployment, and hardware-cost feedback through estimators or measurements on the target device.

Modern NAS research increasingly emphasizes reproducible search spaces and hardware-aware optimization. Benchmark suites such as NAS-Bench-101~\cite{ying2019bench} and NAS-Bench-201~\cite{dong2020bench} highlight the importance of standardized and reusable search spaces, while hardware-aware NAS approaches show that deployment constraints must be considered during architecture search. However, applying these ideas to custom accelerators and embedded platforms often still requires substantial manual engineering effort, especially when different NAS paradigms or target platforms have to be studied within the same experimental setup.

To address this gap, we present \textit{\framework}\footnote{Public on GitHub:~\url{https://github.com/es-ude/elastic-ai.explorer}}, a modular framework for hardware-aware neural architecture exploration on embedded targets. Built on Optuna~\cite{akiba2019optuna}, it introduces a declarative YAML-based search space specification that is dynamically translated into executable model instances during sampling. The specification captures layer-wise, cell-based, and hierarchical search spaces in a common format, providing a reusable and extensible basis for defining different types of NAS search spaces.
Beyond model generation, \textit{\framework} provides a modular hardware backend for embedded deployment and evaluation. Generator components translate sampled architectures into hardware-specific artifacts, interact with compilation toolchains, and enable deployment and benchmarking on target devices. Hardware-cost feedback can then be incorporated into the optimization loop, enabling hardware-aware model selection for concrete embedded platforms. This enables an end-to-end approach from defining the search space to deploying and running an artifact on the hardware target.
The main contributions of this work are:
\vspace*{-1mm}
\begin{enumerate}
    \item We present \textit{\framework}, an end-to-end
    framework for hardware-aware neural architecture exploration
    on resource-constrained embedded devices.
    \item We introduce a declarative and extensible search-space
    specification for systematically defining and generating neural
    architectures.
    \item We provide a modular hardware backend that enables
    deployment and measurement-backed model selection on physical
    target hardware.
\end{enumerate}

The paper is structured as follows:
Section~\ref{sec:relatedwork}
discusses related work in the area of hardware-aware NAS, with a particular focus on existing NAS frameworks. Section~\ref{sec:arch_overview} provides an architectural overview of the \textit{\framework}. Section~\ref{sec:sp}  describes the search space modeling and dynamic candidate model construction in detail, while section~\ref{evaluation} and~\ref{sec:deployment} cover the evaluation API and hardware deployment respectively.
Finally, we conclude the paper with a brief outlook on future work in section~\ref{sec:conclusion}.

%%%%%%%%%%%%%%%%%%%%%%%%%%%%%%%%%%%%%%%%%%%%%%%%%%%%%%%%%%%%%%%
\section{Related Work}\label{sec:relatedwork}
Neural Architecture Search~(NAS) has been widely studied as a means to automate the design of neural network architectures~\cite{elsken2019survey}. By exploring alternative architectural configurations automatically, NAS aims to identify models that perform well for a given task. In embedded settings, this objective must be extended to also account for hardware-specific constraints such as latency, memory footprint, energy consumption, or model size. Hardware-aware NAS addresses this by incorporating such objectives into the search process. Approaches such as MnasNet~\cite{mingxing2018mnasnet}, FBNet~\cite{wu2018fbnet}, ProxylessNAS~\cite{cai2018proxylessnas}, Once-for-All~\cite{cai2019onceforall}, and ChamNet~\cite{dai2018chamnet} have shown that hardware-dependent metrics can guide the search towards architectures that are both accurate and efficient on target platforms.

While these methods demonstrate the relevance of hardware-aware optimization, they are primarily individual NAS methods. Reusable frameworks such as NNI~\cite{nni-2021}, NASLib~\cite{naslib-2020}, and Archai~\cite{archai-2022} provide abstractions for search spaces, optimization strategies, and evaluation workflows, while NNI and Archai further support hardware-related objectives such as latency, memory usage, model size, or computational cost~\cite{nni-2021, archai-2022}. More hardware-oriented frameworks such as AW-NAS~\cite{aw-nas2020} and HANNAH~\cite{hannah-2025} integrate hardware-specific cost models or backend-based evaluation into the search process. AW-NAS is an important example of a modular hardware-aware NAS framework, but is no longer actively maintained. Among these hardware-oriented frameworks, HANNAH is most relevant for comparison, as it combines NAS with hardware-aware evaluation and parameterized graph-based search space modeling.

In contrast, \textit{\framework} focuses on a declarative YAML-based search space specification that represents different types of NAS search spaces within a common format. Combined with its modular hardware backend, this connects search space specification, model generation, deployment, and on-device measurement within a single toolbox. In addition, the search space can include pre-processing components in order to enable end-to-end signal processing for continuous data stream systems.

\section{Architecture Overview}\label{sec:arch_overview}
The \textit{\framework} framework is organized into two main parts, the NAS toolkit and the hardware-aware deployment pipeline.
An overview of the Explorer architecture can be seen in Figure~\ref{fig:overview}.
The NAS toolkit covers search space translation, candidate sampling, model construction, and evaluation, while the hardware-aware deployment pipeline connects generated model instances to target-specific toolchains and benchmarking workflows.

The search process starts from a YAML-based specification that defines the model search space (see Section~\ref{sec:sp}) and, optionally, a sensor-signal pre-processing configuration (see Section~\ref{sec:sppre}). Both components are parsed by the search space translator, which converts the declarative representation into an Optuna-compatible search space. The pre-processing configuration captures signal-processing operations applied before model inference. This allows pre-processing parameters and architecture parameters to be jointly optimized within the same NAS workflow and enables highly optimized end-to-end signal processing pipelines in continuous sensor data stream applications.

During optimization, sampled trial parameters are transformed into an intermediate architectural representation and instantiated dynamically as an executable model instance by the~\texttt{ModelBuilder}. Based on the sampled configuration, the~\texttt{ModelBuilder} constructs the selected operations, infers intermediate tensor shapes, and inserts adapter modules where required to connect incompatible layer types, eliminating the need to manually implement the transition logic for each sampled architecture.

Candidate architectures are assessed through the evaluation API. The API provides a modular interface for performance and cost estimators, which can be used as objectives, soft constraints, or hard constraints during optimization. Details on the estimator interface and the aggregation of optimization criteria are described in Section~\ref{evaluation}. 

After model sampling and evaluation, selected architectures can be passed to one of the generators. These components are responsible for translating executable model instances into target-specific artifacts, interacting with compilation toolchains, and enabling deployment and benchmarking on embedded target platforms. Furthermore, cost estimators can directly leverage these generators to perform hardware-in-the-loop optimization, enabling accurate measurement-based cost evaluation on real target devices.
The generators are described in more detail in Section~\ref{sec:deployment}.

%%%%%%%%%%%%%%%%%%%%%%%%%%%%%%%%%%%%%%%%%%%%%%%%%%%%%%%%%%%%%%%%%%%%%%%%%%%%%%%%%%%%%%%%%%
\section{Search Space Modeling}\label{sec:sp}
\begin{figure}[ht]  \centering
    \hspace*{-6mm}
    \includegraphics[width=0.54\textwidth]{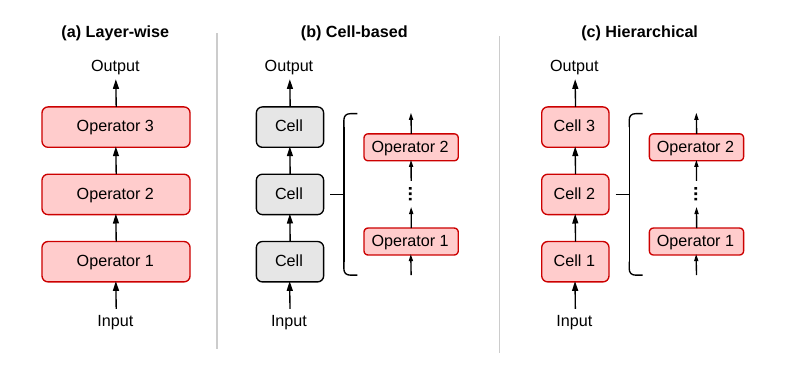}
    \vspace*{-8mm}
    \caption{Different types of NAS search space paradigms: (a) layer-wise - (b) cell-based - (c) hierarchical. Red components are considered during NAS. Grey components are decided manually. Image adapted from~\cite{benmeziane2021comprehensive}.}
    \label{fig:searchspaceParadigms}
\end{figure}

NAS methods differ fundamentally in how they define and constrain the underlying architecture search space. As illustrated in Figure~\ref{fig:searchspaceParadigms}, three dominant paradigms can be identified in the literature: layer-wise~\cite{wu2018fbnet}\cite{baker2017}, cell-based~\cite{pham2018}, and hierarchical search spaces~\cite{mingxing2018mnasnet}\cite{schrodi2023}. These paradigms vary in granularity, parameter sharing, and structural reuse, and therefore require different modeling abstractions. 
%%%

Layer-wise search spaces define architectures as sequences of independently sampled layers, enabling maximum flexibility but resulting in large search spaces. In contrast, cell-based approaches search for reusable computational motifs called cells that are stacked to form deep architectures, significantly reducing search complexity while promoting transferability. Hierarchical search spaces extend this idea by introducing multi-level structure, combining macro-level architectural decisions with fine-grained intra-block search, thereby enabling structured and constrained design spaces.

To support all of these paradigms within a unified optimization framework, the explorer introduces a declarative search space representation that decouples architecture specification from implementation. This representation is expressed in a YAML-based domain-specific language (DSL), which is automatically translated into an Optuna-compatible search space and dynamically instantiated as executable model instances during optimization.
%%%%%%%%%%%%%%%%%%%%%%%%%%%%%%%%%%%%%%%%%%%%%%%%%%%%%%%%%%%%%%%%%%%%%%%%%%%%%%%%%%%%%%%%%%
\subsection{Search Space Structure}\label{sec:sss}
The core abstraction of the DSL is a block-based representation of the network. A search space is defined as a sequence of blocks, where each block corresponds to a contiguous portion of the final architecture. The top-level structure of the yaml file is shown in Listing~\ref{lst:Top-levelSP} and a minimal block specification in Listing~\ref{lst:block}.
%%%
\definecolor{placeholder}{RGB}{180,40,40}
%\begin{listing}
%\begin{minted}[fontsize=\small, autogobble=true , bgcolor=bg, breaklines]{yaml}
\begin{lstlisting}[caption={Top-Level Syntax of the search space DSL. Values in angle brackets are set by user.},captionpos=b, label={lst:Top-levelSP}]
input: <INPUT_SHAPE>
output: <OUTPUT_SHAPE>

sequence:
  - block: <UNIQUE_BLOCK_NAME>
    ...
  - ...

default_op_params:      # optional
  <OP_NAME>:     
    <PARAMETER_NAME>: <VALUE | CHOICES>
        ...
    ...
composites:             # optional
  <COMPOSITE_NAME>:
    sequence:
      - block: <UNIQUE_BLOCK_NAME>
        ...
\end{lstlisting}
%\end{minted}
%  sequence:             
%   - block: <unique_name>         
%       ...               
%   - block: <unique_name>
%       ...
      
% default_op_params:     
%     primitive_op_1:
%       param_1: [...]    
%       param_2: [...]
%     primitive_op_2:
%       param_1: [..]
      
% composites:
%     <unique_operator_name>:
%       sequence:
%         - block: <unique_name> 
%             ...

%\begin{listing}
%\begin{minted}[fontsize=\small, autogobble=true , bgcolor=bg, breaklines]{yaml}

Each block is assigned a unique identifier and a set of~\verb|op_candidates|, which defines the set of operations that may be sampled by the NAS algorithm. If an operation is suggested for a block, its associated mandatory parameters must be defined either locally within the block or globally via the~\verb|default_op_params| section. The global parameter section acts as a fallback mechanism, allowing shared definitions across blocks and reducing redundancy in the search space specification.
%%%
\begin{Code}
\begin{lstlisting}[caption={Search space block Syntax. Values in angle brackets are set by user.}, captionpos=b, label={lst:block}]
- block: <UNIQUE_BLOCK_NAME>

  op_candidates: <OP_NAME>| LIST[<OP_NAME>]

  type_repeat:                # optional
    type: <REPEAT_MODE>
    depth: <INT | CHOICES[INT]>  # optional
    ref_block: <REF_BLOCK_NAME> #optional

  <OP_NAME>:
    <PARAMETER_NAME>: <VALUE | CHOICES>
    ...
  <OP_NAME_2>:
    ...
\end{lstlisting}
\end{Code}
%\end{minted}
%   - block:  "unique_id"
%     op_candidates:  ["op_1", "op_2"]
%     type_repeat:
%       type: <repeat_type>
%       depth: [1, 2]
%     op1:
%       op_param_1: [...]
%       op_param_2: [...]
%     op2:
%       ...

%%%%%%%%%%%%%%%%%%%%%%%%%%%%%%%%%%%%%%%%%%%%%%%%%%%%%%%%%%%%%%%%%%%%%%%%%%%%%%%%%%%%%%%%%%
%\begin{listing}[t]
%\begin{minted}[fontsize=\small, autogobble=true, bgcolor=bg, breaklines]{yaml}
\begin{Code}
\begin{lstlisting}[caption={Example YAML search space for a one-dimensional convolutional classifier.}, captionpos=b , label={lst:yamlExample}]
input: [4, 1250]
output: 6
sequence:
  - block: "features"
    op_candidates: "conv-block"
    type_repeat:
      type: "vary_all"
      depth: [1, 2, 3, 4, 5, 6]
  - block: "head"
    op_candidates: "linear"
    linear:
      width: [32, 64, 128]
default_op_params:
  conv1d:
    kernel_size: [3, 5]
    out_channels: [8, 16]
composites:
  conv-block:
    sequence:
      - block: "conv"
        op_candidates: "conv1d"
      - block: "pool"
        op_candidates: ["maxpool", "identity"]
\end{lstlisting}
\end{Code}
%\end{minted}

\subsection{Block Repetition and Composites}
Beyond individual block definitions, the DSL introduces a flexible repetition mechanism that determines how operations and parameters are shared or varied across layers within a block. If no repetition rule is specified, the block is treated as a single layer.

The repetition behavior is controlled via the \verb|type_repeat| field, which defines both the repetition strategy and the depth of repetition. Table~\ref{tab:repeat_type} summarizes the supported modes.
Furthermore, to further support hierarchical and cell-based design patterns, the DSL introduces composite operations. The composite operations can be defined in the top-level \verb|composites| section of the search space seen in Listing~\ref{lst:Top-levelSP}. 
A composite defines a reusable sub search space consisting of a sequence of blocks that can be embedded into other architectures. This mechanism enables modular construction of complex search spaces and promotes reuse of architectural components across different experiments. 
Composites are particularly useful for encoding frequently occurring patterns such as convolution-activation-pooling pipelines, which can then be treated as atomic operations within higher-level search spaces.

Combining the different repetition modes with the composite operations allows for modeling the different search space paradigms from Figure~\ref{fig:searchspaceParadigms}. 
\begin{table*}[!ht]
    \centering
    \caption{Different repeat options for Block operator sampling.}
    \begin{tabularx}{\textwidth}{l X | X}
    \toprule
        type & Description & Mandatory fields \\ 
        \midrule
        repeat\_op & Repeat the same operation for the chosen depth. Searched parameters for each layer may change. & \multirow[c]{3}{*}{ Depth of type\_repeat must also be set.}\\ \cmidrule{1-2}
        repeat\_params & Only search the operations and its parameters once per block. Reuse sampled params for every layer in the block. &  \\ 
        \cmidrule{1-2}
        vary\_all & Vary operations and parameters for each layer in block & \\ 
        \midrule
        repeat\_block & Repeat an already defined block Input can vary depending on previous block & reference\_block must be set with unique block identifier of block to be repeated \\ 
        \bottomrule
    \end{tabularx}
    \label{tab:repeat_type}
\end{table*}

\begin{itemize}
    \item \textbf{Layer-wise search spaces} are represented by \texttt{vary\_all}, where both operations and parameters are independently sampled for each layer, resulting in maximum flexibility and minimal sharing.
    
    \item \textbf{Cell-based search spaces} can be expressed through \texttt{repeat\_params}, \texttt{repeat\_ops} or \texttt{repeat\_block}. Composite operations further facilitate stacking of more complex cells. These modes enforce structural reuse by repeating either the sampled operation or both operation and parameters across layers, corresponding to reusable architectural motifs.
    
    \item \textbf{Hierarchical search spaces} are modeled through nested composite operations and combinations of all the available repetition mechanisms, enabling multi-level architectural composition and reuse of higher-level structural building blocks.
\end{itemize}

Listing~\ref{lst:yamlExample} illustrates a concrete YAML search space for a one dimensional convolutional classifier. The example defines an input shape, an output dimension, a repeated feature extraction block, and a classification head. The feature extraction block references the reusable \texttt{conv-block}, which itself consists of a sequence of primitive operations. Searchable parameters are either defined locally within a block or globally through the \texttt{default\_op\_params} section.
%%%%%%%%%%%%%%%%%%%%%%%%%%%%%%%%%%%%%%%%%%%%%%%%%%%%%%%%%%%%%%%%%%%%%%%%%%%%%%%%%%%%%%%%%%
\subsection{Dynamic Instantiation and Model Construction}
During optimization, sampled configurations are automatically translated into executable PyTorch models. The framework instantiates modules only after Optuna selects parameter values. This allows conditional dependencies and recursive architectural structures. The framework dynamically constructs neural networks at runtime by instantiating selected operations and inserting required adapter modules between incompatible layer types.
This mechanism enables heterogeneous architectures to be sampled without requiring manual implementation of shape transformations or structural compatibility logic. As a result, the same search space definition can support convolutional, recurrent, and fully connected architectures.
%%%%%%%%%%%%%%%%%%%%%%%%%%%%%%%%%%%%%%%%%%%%%%%%%%%%%%%%%%%%%%%%%%%%%%%%%%%%%%%%%%%%%%%%%%
\subsection{Extensibility via Operation Registration}
The framework is designed to be fully extensible with minimal implementation effort. New operations can be introduced by implementing a \verb|LayerBuilder| Interface and registering it via a decorator. The Interface and an example decorator registration can be seen in Listing~\ref{lst:layerBuilderInterface}.
%\begin{listing}
%\begin{minted}[fontsize=\small, autogobble=true, bgcolor=bg, breaklines]{python}
\begin{lstlisting}[caption={Layer Builder Interface and example registration of a linear layer.}, captionpos=b, label={lst:layerBuilderInterface}]
class LayerBuilder(ABC):

    @abstractmethod
    def build_layer(self, input_shape, search_parameters: dict):
        pass

    @abstractmethod
    def get_last_layer(self, input_shape, search_parameters: dict, output_shape):
        pass

@register_layer("linear")
class LinearLayer(LayerBuilder):
        ...
\end{lstlisting}
%\end{minted}

Each added operation must define how it is constructed from sampled parameters and how its output shape is computed. 
Because operations are sometimes treated differently if they are the last layers in the network, the interface makes this distinction.
After registering a specific layer with the \verb|@register_layer("op_name")| decorator, the operation can be used in the yaml under the same name.

In cases where transitions between operations require non-trivial transformations, adapter modules can be defined to handle shape conversion and structural compatibility. These adapters are registered in a transition registry that maps pairs of operation types to corresponding transformation modules.

This plugin-based design allows researchers to integrate custom layers or hardware-specific primitives without modifying the core NAS engine, enabling seamless adaptation to new hardware platforms and accelerator designs.

%%%%%%%%%%%%%%%%%%%%%%%%%%%%%%%%%%%%%%%%%%%%%%%%%%%%%%%%%%%%%%%%%%%%%%%%%%%%%%%%%%%%%%%%%%
\subsection{Pre-processing Design Space}\label{sec:sppre}
Following the same modeling principle as for neural architectures, the \textit{\framework} also allows pre-processing pipelines to be included in the search process. By considering the pre-processing in the NAS, this allows to transform a continuous data streaming process into an event-based approach for running the deep learning model optimally and to enable an end-to-end optimization of the whole sensor data processing pipeline~\cite{Buron2023}. In its current implementation, the searchable configuration is primarily tailored to time-series and sensor-signal data and supports five configurable pre-processing operations: filtering, downsampling, sequential windowing, event-based windowing and normalization.

During optimization, the selected operations and parameters can be sampled as part of the same trial as the architecture configuration to instantiate an executable pre-processing pipeline. The sampled pipeline transforms the input data before it is passed to the corresponding model, enabling joint exploration of pre-processing and architecture choices within a single search process.
\section{Evaluation API}\label{evaluation}
The evaluation API defines how sampled architectures are assessed during the search process and separates metric computation from the core NAS workflow.
Performance estimators evaluate task-specific objectives such as classification accuracy or reconstruction loss. Cost estimators provide hardware-related metrics including parameter count, FLOPs, memory consumption, and latency estimates. Since evaluators are implemented independently of the NAS workflow, custom optimization objectives can be integrated without modifications to the search infrastructure.

Cost and performance estimators can either be used directly as Optuna objectives, including Optuna's native multi-objective optimization algorithms, or be registered as an~\texttt{OptimizationCriteria} within the framework. Optimization criteria can be objectives, soft constraints, or hard constraints.
During each optimization trial, registered criteria are evaluated sequentially. Hard constraints are evaluated first and may terminate invalid configurations early if predefined limits are violated. Afterwards, objective functions and soft constraints are evaluated on the remaining valid architectures. This staged evaluation reduces unnecessary computation for architectures that already fail mandatory deployment or resource requirements.

By default, objective values and soft constraints are scalarized by means of a weighted sum, requiring user-defined weighting factors for the individual criteria. In addition to the default scalarization strategy, custom optimization aggregation functions can be injected. This enables optimization across heterogeneous objectives such as task accuracy, latency, parameter count, or memory consumption into a single unified optimization score. Measurement-based cost values, such as hardware latency, can be provided by the hardware-aware deployment pipeline described in~ Section~\ref{sec:deployment}.

\section{Hardware-Aware Deployment Pipeline}\label{sec:deployment}
A central feature of~\textit{\framework} is its integrated deployment and benchmarking pipeline. The framework contains a generator system that transforms sampled architectures into deployable implementations for embedded platforms in order to evaluate their performance. The red box in Figure~\ref{fig:overview} shows the generator pipeline, which consists of model builders, compilers, host interfaces, and hardware managers. In order to enable reproducible results across different systems and OS, generated artifacts can be compiled using Docker-based cross-compilation toolchains. This significantly reduces setup complexity for embedded deployment.

In general, the framework supports hardware backends for translating the generated model instances into hardware-specific executions for different platforms. It supports embedded hardware like Raspberry Pi 4/5 using~\textit{torchscript}, microcontrollers like Raspberry Pi Pico 1/2 (W) using~\textit{LiteRt} and FPGAs like elasticAI.ENV5 using the \textit{elasticAI.creator}~\cite{chao2023creator}.

This generator pipeline can be used in two modes. In the first mode, the best model from NAS is translated into hardware executions in dependency of the toolchain for the hardware target. The resulting artifact can be deployed on hardware and latency and model performance can be evaluated. In the second mode, the generator pipeline is used during the NAS process in a hardware-in-the-loop setting. Here, the framework translates the candidates to a format suitable for the target hardware and extracts on device cost metrics such as latency automatically instead of relying solely on analytical assumptions of an analytical cost estimator. Benchmarked values are then reintegrated into the NAS optimization loop.

In addition, a generator can define a reflection API which can be used by the model builder to take the specific hardware's and generator's capabilities into consideration. This guarantees that only supported operations and implementations will be accessed during the search process. Furthermore, this allows for generator specific layer implementations to be used instead of the default one's.
\section{Conclusion and Outlook}\label{sec:conclusion}
\textit{\framework} extends the Optuna ecosystem with a flexible and deployment-oriented framework for hardware-aware NAS. Through YAML-based search space specifications and dynamic model generation, the framework enables expressive NAS formulations while maintaining extensibility. Its integrated generator pipeline, Docker-based cross-compilation support, and hardware-in-the-loop benchmarking capabilities provide a practical foundation for embedded AI research and deployment.
Future work will extend~\textit{\framework} with support for quantization- and pruning-aware NAS to better address the resource constraints of embedded target platforms. There, we want to enhance the hardware backend with support for FPGAs and custom SoCs. In addition, we plan to investigate supernet-based and one-shot NAS approaches to improve the efficiency of the search process.

%%%%%%%%%%%%%%%%%%%%%%%%%%%%%%%%%%%%%%%%%%%%%%%%%%%%%%%%%%%%%%%%%%%
%% LITERATURE
%%%%%%%%%%%%%%%%%%%%%%%%%%%%%%%%%%%%%%%%%%%%%%%%%%%%%%%%%%%%%%%%%%%
\bibliographystyle{IEEEtran}
\bibliography{bibliography}

@inproceedings{akiba2019optuna,
    title = {{Optuna: A next-generation hyperparameter optimization framework}},
    author = {Akiba, Takuya and Sano, Shotaro and Yanase, Toshihiko and Ohta, Takeru and Koyama, Masanori},
    booktitle = {Proceedings of the 25th ACM Int. Conf. on Knowledge Discovery \& Data Mining (SIGKDD)},
    pages = {2623--2631},
    year = {2019},
    url       = {https://doi.org/10.1145/3292500.3330701}
}

@article{dong2020bench,
    title = {{Nas-bench-201: Extending the Scope of Reproducible Neural Architecture Search}},
    author = {Dong, Xuanyi and Yang, Yi},
    journal = {arXiv preprint arXiv:2001.00326},
    year = {2020},
    url       = {https://openreview.net/forum?id=HJxyZkBKDr},
}

@inproceedings{ying2019bench,
    title = {{Nas-bench-101: Towards Reproducible Neural Architecture Search}},
    author = {Ying, Chris and Klein, Aaron and Christiansen, Eric and Real, Esteban and Murphy, Kevin and Hutter, Frank},
    booktitle = {Int. Conf. on Machine Learning},
    pages = {7105--7114},
    year = {2019},
    organization = {PMLR},
    url       = {https://proceedings.mlr.press/v97/ying19a.html},
}

@article{benmeziane2021comprehensive,
    title = {{A Comprehensive Survey on Hardware-Aware Neural Architecture Search}},
    author = {Benmeziane, Hadjer and Maghraoui, Kaoutar El and Ouarnoughi, Hamza and Niar, Smail and Wistuba, Martin and Wang, Naigang},
    journal = {arXiv preprint arXiv:2101.09336},
    year = {2021},
    url     = {https://arxiv.org/abs/2101.09336},
}

@article{Buron2023,
    author    = {Leo Buron and Andreas Erbsl{\"o}h and Zia Ur-Rehman and Christian Klaes and Karsten Seidl and Gregor Schiele},
    title     = {{Deep.Neural.Signal.Pre-Processor -- Towards Development of AI-enhanced End-to-End BCIs}},
    journal   = {Current Directions in Biomedical Engineering},
    volume    = {9},
    number    = {1},
    pages     = {471--474},
    year      = {2023},
    publisher = {Walter de Gruyter GmbH},
    doi       = {10.1515/cdbme-2023-1118},
    url       = {https://doi.org/10.1515/cdbme-2023-1118}
}

@article{elsken2019survey,
  title   = {{Neural Architecture Search: A Survey}},
  author  = {Thomas Elsken and Jan Hendrik Metzen and Frank Hutter},
  journal = {Journal of Machine Learning Research},
  volume  = {20},
  number  = {55},
  pages   = {1--21},
  year    = {2019},
  url     = {https://www.jmlr.org/papers/v20/18-598.html}
}

@article{mingxing2018mnasnet,
    author       = {Mingxing Tan and Bo Chen and Ruoming Pang and Vijay Vasudevan and Quoc V. Le},
    title        = {{MnasNet: Platform-Aware Neural Architecture Search for Mobile}},
    journal      = {CoRR},
    volume       = {abs/1807.11626},
    year         = {2018},
    url          = {http://arxiv.org/abs/1807.11626},
    eprinttype   = {arXiv},
    eprint       = {1807.11626},
}

@article{wu2018fbnet,
    author       = {Bichen Wu and Xiaoliang Dai and Peizhao Zhang and Yanghan Wang and Fei Sun and Yiming Wu and Yuandong Tian and Peter Vajda and Yangqing Jia and Kurt Keutzer},
    title        = {{FBNet: Hardware-Aware Efficient ConvNet Design via Differentiable Neural Architecture Search}},
    journal      = {CoRR},
    volume       = {abs/1812.03443},
    year         = {2018},
    url          = {http://arxiv.org/abs/1812.03443},
    eprinttype   = {arXiv},
    eprint       = {1812.03443},
}

@article{cai2018proxylessnas,
    author       = {Han Cai and Ligeng Zhu and Song Han},
    title        = {{ProxylessNAS: Direct Neural Architecture Search on Target Task and Hardware}},
    journal      = {CoRR},
    volume       = {abs/1812.00332},
    year         = {2018},
    url          = {http://arxiv.org/abs/1812.00332},
    eprinttype   = {arXiv},
    eprint       = {1812.00332},
}

@article{cai2019onceforall,
    author       = {Han Cai and Chuang Gan and Song Han},
    title        = {{Once for All: Train One Network and Specialize it for Efficient Deployment}},
    journal      = {CoRR},
    volume       = {abs/1908.09791},
    year         = {2019},
    url          = {http://arxiv.org/abs/1908.09791},
    eprinttype   = {arXiv},
    eprint       = {1908.09791},
}

@article{dai2018chamnet,
    author       = {Xiaoliang Dai and Peizhao Zhang and Bichen Wu and Hongxu Yin and Fei Sun and Yanghan Wang and Marat Dukhan and Yunqing Hu and Yiming Wu and Yangqing Jia and Peter Vajda and Matt Uyttendaele and Niraj K. Jha},
    title        = {{ChamNet: Towards Efficient Network Design through Platform-Aware Model Adaptation}},
    journal      = {CoRR},
    volume       = {abs/1812.08934},
    year         = {2018},
    url          = {http://arxiv.org/abs/1812.08934},
    eprinttype   = {arXiv},
    eprint       = {1812.08934},
}

@misc{naslib-2020, 
    title = {{NASLib: A Modular and Flexible Neural Architecture Search Library}}, 
    author = {Ruchte, Michael and Zela, Arber and Siems, Julien and Grabocka, Josif and Hutter, Frank}, 
    year = {2020}, 
    publisher = {GitHub}, 
    howpublished = {\url{https://github.com/automl/NASLib}} 
}

@misc{archai-2022,
    title = {{Archai: Platform for Neural Architecture Search}},
    url = {https://www.microsoft.com/en-us/research/project/archai-platform-for-neural-architecture-search},
    journal = {Microsoft Research},
    year = {2022},
    month = {Jul}
}

@software{nni-2021,
   author = {{Microsoft}},
   month = {1},
   title = {{Neural Network Intelligence}},
   url = {https://github.com/microsoft/nni},
   version = {2.0},
   year = {2021}
}

@article{aw-nas2020,
  author       = {Xuefei Ning and Changcheng Tang and Wenshuo Li and Songyi Yang and Tianchen Zhao and Niansong Zhang and Tianyi Lu and Shuang Liang and Huazhong Yang and Yu Wang},
  title        = {{aw{\_}nas: {A} Modularized and Extensible {NAS} framework}},
  journal      = {CoRR},
  volume       = {abs/2012.10388},
  year         = {2020},
  url          = {https://arxiv.org/abs/2012.10388},
  eprinttype   = {arXiv},
  eprint       = {2012.10388},
}

@INPROCEEDINGS{hannah-2025,
    author = {Christoph, Gerum and Adrian, Frischknecht and Tobias, Hald and Bernardo, Paul Palomero and Lübeck, Konstantin and Oliver, Bringmann},
    booktitle = {25th Euromicro Conference on Digital System Design (DSD)}, 
    title = {{Hardware Accelerator and Neural Network Co-Optimization for Ultra-Low-Power Audio Processing Devices}}, 
    year = {2022},
    volume = {},
    number = {},
    pages = {365-369},
    doi = {10.1109/DSD57027.2022.00056},
    url       = {https://doi.org/10.1109/DSD57027.2022.00056}
}

@misc{baker2017,
    title = {{Designing Neural Network Architectures using Reinforcement Learning}}, 
    author = {Bowen Baker and Otkrist Gupta and Nikhil Naik and Ramesh Raskar},
    year = {2017},
    eprint = {1611.02167},
    archivePrefix = {arXiv},
    primaryClass = {cs.LG},
    url = {https://arxiv.org/abs/1611.02167}, 
}

@misc{pham2018,
    title = {{Efficient Neural Architecture Search via Parameter Sharing}}, 
    author = {Hieu Pham and Melody Y. Guan and Barret Zoph and Quoc V. Le and Jeff Dean},
    year = {2018},
    eprint = {1802.03268},
    archivePrefix = {arXiv},
    primaryClass = {cs.LG},
    url = {https://arxiv.org/abs/1802.03268}, 
}

@misc{schrodi2023,
    title = {{Construction of Hierarchical Neural Architecture Search Spaces based on Context-free Grammars}}, 
    author = {Simon Schrodi and Danny Stoll and Binxin Ru and Rhea Sukthanker and Thomas Brox and Frank Hutter},
    year = {2023},
    eprint = {2211.01842},
    archivePrefix = {arXiv},
    primaryClass = {cs.LG},
    url = {https://arxiv.org/abs/2211.01842}, 
}

@inproceedings{chao2023creator,
    author = {Qian, Chao and Einhaus, Lukas and Schiele, Gregor},
    title = {{ElasticAI-Creator: Optimizing Neural Networks for Time-Series-Analysis for on-Device Machine Learning in IoT Systems}},
    year = {2023},
    url = {https://doi.org/10.1145/3560905.3568296},
    doi = {10.1145/3560905.3568296},
    booktitle = {Proceedings of the 20th ACM Conf. on Embedded Networked Sensor Systems (SenSys '22)},
    pages = {941–946},
    numpages = {6},
    location = {Boston, Massachusetts}
}

\end{document}